\documentclass[3p,times,twocolumn]{elsarticle}

\usepackage{ecrc}
\usepackage{lscape}

\volume{00}

\firstpage{1}

\journalname{Nuclear Physics B Proceedings Supplement}

\runauth{}

\jid{nuphbp}

\jnltitlelogo{Nuclear Physics B Proceedings Supplement}

\usepackage{amssymb}

\usepackage[figuresright]{rotating}

\begin{document}

\begin{frontmatter}



\dochead{}

\title{Results of the material screening program of the NEXT experiment}


\author[lsc,zgz]{T.~Dafni\corref{cor1}}
\author[ific]{V.~\'Alvarez}
\author[lsc]{I.~Bandac}
\author[lsc,padova]{A.~Bettini}
\author[coimbra]{F.I.G.M.~Borges}
\author[colombia]{M.~Camargo}
\author[ific]{S.~C\'arcel}
\author[lsc,zgz]{S.~Cebri\'an}
\author[ific]{A.~Cervera}
\author[coimbra]{C.A.N.~Conde}
\author[ific]{J.~D\'iaz}
\author[upv]{R.~Esteve}
\author[coimbra]{L.M.P.~Fernandes}
\author[ciemat]{M.~Fern\'andez}
\author[ific]{P.~Ferrario}
\author[aveiro]{A.L.~Ferreira}
\author[coimbra]{E.D.C.~Freitas}
\author[berkeley]{V.M.~Gehman}
\author[berkeley]{A.~Goldschmidt}
\author[lsc,zgz]{H.~G\'omez}
\author[ific]{J.J.~G\'omez-Cadenas}
\author[lsc,zgz]{D.~Gonz\'alez-D\'iaz}
\author[colombia]{R.M.~Guti\'errez}
\author[iowa]{J.~Hauptman}
\author[santiago]{J.A.~Hernando Morata}
\author[lsc,zgz]{D.C.~Herrera}
\author[lsc,zgz]{F.J.~Iguaz}
\author[lsc,zgz]{I.G.~Irastorza}
\author[uam]{L.~Labarga}
\author[ific]{A.~Laing}
\author[ific]{I.~Liubarsky}
\author[ific]{D.~Lorca}
\author[colombia]{M.~Losada}
\author[lsc,zgz]{G.~Luz\'on}
\author[upv]{A.~Mar\'i}
\author[ific]{J.~Mart\'in-Albo}
\author[ific]{A.~Mart\'inez}
\author[santiago]{G.~Mart\'inez-Lema}
\author[berkeley]{T.~Miller}
\author[ific]{F.~Monrabal}
\author[ific]{M.~Monserrate}
\author[coimbra]{C.M.B.~Monteiro}
\author[upv]{F.J.~Mora}
\author[aveiro]{L.M. Moutinho}
\author[ific]{J.~Mu\~noz Vidal}
\author[ific]{M.~Nebot-Guinot}
\author[berkeley]{D.~Nygren}
\author[berkeley]{C.A.B.~Oliveira}
\author[ift]{J.~P\'erez}
\author[upv2]{J.L.~P\'erez Aparicio}
\author[berkeley]{J.~Renner}
\author[girona]{L.~Ripoll}
\author[lsc,zgz]{A.~Rodr\'iguez}
\author[ific]{J.~Rodr\'iguez}
\author[coimbra]{F.P.~Santos}
\author[coimbra]{J.M.F.~dos Santos}
\author[lsc,zgz]{L.~Segui}
\author[ific]{L.~Serra}
\author[berkeley]{D.~Shuman}
\author[ific]{A. Sim\'on}
\author[texas]{C.~Sofka}
\author[ific]{M.~Sorel}
\author[upv]{J.F.~Toledo}
\author[girona]{J.~Torrent}
\author[dubna]{Z.~Tsamalaidze}
\author[aveiro]{J.F.C.A.~Veloso}
\author[lsc,zgz]{J.A.~Villar}
\author[texas]{R.C.~Webb}
\author[texas]{J.T.~White}
\author[ific]{N.~Yahlali}

\cortext[cor1]{Attending speaker}
\address[lsc]{Laboratorio Subterráneo de Canfranc, 22880 Canfranc Estación, Huesca, Spain}
\address[zgz]{Laboratorio de F\'isica Nuclear y Astropart\'iculas, Universidad de Zaragoza, 50009 Zaragoza, Spain}
\address[ific]{Instituto de F\'isica Corpuscular (IFIC), CSIC \& Universitat de Val\`encia, 46980 Paterna, Valencia, Spain}
\address[padova]{Padua University and INFN Section, Dipartimento di Fisca G. Galilei, 35131 Padova, Italy} 
\address[coimbra]{Departamento de Fisica, Universidade de Coimbra, 3004-516 Coimbra, Portugal}
\address[colombia]{Centro de Investigaciones en Ciencias B\'asicas y Aplicadas, Universidad Antonio Nari\~no, Bogot\'a, Colombia}
\address[upv]{Instituto de Instrumentaci\'on para Imagen Molecular (I3M), U. Polit\`ecnica de Val\`encia, 46022 Valencia, Spain}
\address[ciemat]{Centro de Investigaciones Energ\'eticas, Medioambientales y Tecnol\'ogicas (CIEMAT), 28040 Madrid, Spain} 
\address[aveiro]{Institute of Nanostructures, Nanomodelling and Nanofabrication (i3N), U. de Aveiro, 3810-193 Aveiro, Portugal}
\address[berkeley]{Lawrence Berkeley National Laboratory (LBNL), Berkeley, California 94720, USA}
\address[iowa]{Department of Physics and Astronomy, Iowa State University, Ames, Iowa 50011-3160, USA}
\address[santiago]{Instituto Gallego de F\'isica de Altas Energ\'ias (IGFAE), U. de Santiago de Compostela, 15782 Santiago de Compostela, Spain}
\address[uam]{Departamento de F\'isica Te\'orica, Universidad Aut\'onoma de Madrid, 28049 Madrid, Spain}
\address[ift]{Instituto de F\'isica Te\'orica (IFT), UAM/CSIC, 28049 Madrid, Spain}
\address[upv2]{Dpto.\ de Mec\'anica de Medios Continuos y Teor\'ia de Estructuras, U. Polit\`ecnica de Val\`encia, 46071 Valencia, Spain}
\address[girona]{Escola Polit\`ecnica Superior, Universitat de Girona, 17071 Girona, Spain}
\address[texas]{Department of Physics and Astronomy, Texas A\&M University,Texas 77843-4242, USA}
\address[dubna]{Joint Institute for Nuclear Research (JINR), 141980 Dubna, Russia}

\begin{abstract}
The "Neutrino Experiment with a Xenon TPC" (NEXT), intended to investigate neutrinoless double beta decay, requires extremely low background levels. An extensive material screening and selection process to assess the radioactivity of components is underway combining several techniques, including germanium $\gamma$-ray spectrometry performed at the Canfranc Underground Laboratory; recent results of this material screening program are presented here.
\end{abstract}


\end{frontmatter}


\section{Introduction}
\label{}
The NEXT experiment \cite{next} will operate at the Laboratorio Subterráneo de Canfranc (LSC), Spain, a high-pressure xenon time projection chamber (TPC) to search for neutrinoless double beta decay events of $^{136}$Xe using 100~kg of enriched xenon at 90\%. As in any experiment investigating rare event phenomena, ultra-low background conditions are a must and materials used in the set-up have to be carefully selected. A thorough material screening program was undertaken to evaluate the radioactivity of all the relevant components of NEXT \cite{jinstrp,aiprp}; new results are presented here.

This screening program is mainly based on germanium $\gamma$-ray spectrometry using ultra-low background detectors from the Radiopurity Service of LSC (in particular, those named GeOroel, GeAnayet, GeTobazo, GeLatuca) operated at a depth of 2450 m.w.e.. Detectors are p-type close-end coaxial 2.2-kg High Purity germanium detectors, from Canberra France.
For the measurements presented here, shield consisted of 5 cm of copper in the inner part surrounded by 20 cm of low activity lead, with nitrogen flush to avoid airborne radon intrusion. Detection efficiency is estimated for each sample by GEANT4 simulation. Complementing germanium spectrometry results, measurements based on Glow Discharge Mass Spectrometry (GDMS) and Inductively Coupled Plasma Mass Spectrometry (ICPMS) have been also carried out. GDMS is performed by Evans Analytical Group in France, providing concentrations of U, Th and K. An ICPMS measurement was made at CIEMAT (Unidad de Espectrometria de Masas) in Spain.

\section{Results}
\label{}

Materials analyzed deal with the shielding, pressure vessel, field cage and electroluminescence (EL) components and the energy and tracking readout planes. Results obtained after those presented in \cite{jinstrp,aiprp} are summarized in table \ref{tr} and described in the following; for germanium measurements, reported errors correspond to 1$\sigma$ uncertainties and upper limits are given at 95\% C.L.. Uncertainties for GDMS results are typically of 20\%.

Lead and copper from different suppliers to be used as shielding were studied \cite{jinstrp,aiprp}. Finally, refurbished lead from the OPERA experiment with 80 Bq/kg of $^{210}$Pb will be used for external shielding (\#1-2) and CuA1 (or ETP) copper will be used for inner shield (\#3-4). For the pressure vessel, several samples of 316Ti Stainless Steel were initially screened with germanium detectors: 10-mm-thick for body, 15-mm-thick for end-caps, 50-mm-thick for flanges. Now, complementary results have been obtained from GDMS analysis (\#5-7).

Concerning the field cage and EL region, several types of plastics \cite{jinstrp,aiprp} and High Density Poliethylene (PE500) for field cage (\#8) have been screened; HD polyethylene has been analyzed also by ICPMS (\#9). In addition, results for silver epoxy (CW2400) (\#10) and ETP copper for field cage, in rod (\#11) and sheet (\#12), have been obtained.

The tracking readout in NEXT is based on SiPMs in kapton Printed Circuit Boards (PCB). PCB boards (made of cuflon \cite{jinstrp,aiprp} or kapton and copper (\#13)) and different electronic components (capacitors, resistors, connectors, solder paste \cite{jinstrp}, NTC temperature sensors (\#14) and blue LEDs (\#15)) have been screened with germanium detectors. Plexiglas sheets which could be placed in front of boards have been also considered (\#16). At the opposite side of the vessel, the energy readout is based on photomultipliers (PMTs); 34 (out of 60) Hamamatsu R11410-10 PMTs have been already screened in 3-unit groups (\#17) showing equivalent activity. Shappire windows \cite{aiprp} and copper have been studied: CuA1 (or ETP) for PMT cans (\#3-4) and CuC1 (or OF) for plates (\#18). Several components for PMT bases have been also analyzed: capacitors (\#19), resistors (\#20), pin receptacles (\#21) and thermal epoxy (\#22).

In summary, complementary activity measurements based on ICPMS, GDMS and germanium spectrometry performed at LSC have been carried out to help both in the design of the set-up and in the construction of the background model of the NEXT experiment. Radiopure enough samples of copper for inner shielding, stainless steel for pressure vessel and polyethylene for field cage have been found: expected contributions from $^{214}$Bi+$^{208}$Tl at the region of interest are 9.7, 2.9 and 9.4 10$^{-5}$ keV$^{-1}$ kg$^{-1}$ y$^{-1}$ respectively. An extensive work has been carried out, but the screening program is still going on and SiPMs and shielding structure components are now under analysis.



\section*{Acnowledgments} The NEXT Collaboration acknowledges funding support from: the MINECO of Spain under Grants CONSOLIDER-Ingenio 2010 CSD2008-0037 (CUP), FPA2009-13697-C04-04, and FIS2012-37947-C04; the Director, Office of Science, Office of Basic Energy Sciences of the US DoE under Contract no. DE-AC02-05CH11231; and the Portuguese FCT and FEDER through the program COMPETE, Projects PTDC/FIS/103860/2008 and PTDC/FIS/112272/2009.



\begin{thebibliography}{00}
\bibitem{next} J. J. Gomez Cadenas et al, Advances in High Energy Physics vol. 2014 907067.
\bibitem{jinstrp} V. Alvarez et al, JINST 8 (2013) T01002.
\bibitem{aiprp} V. Alvarez et al, AIP Conf. Proc. 1549 (2013) 46.

\end{thebibliography}



\footnotesize
\begin{landscape}
\begin{table}
\begin{tabular}{p{0.1cm}p{2.4cm}p{2.3cm}p{1.2cm}p{1cm}p{1.5cm}p{1.5cm}p{1.5cm}p{1.5cm}p{0.8cm}p{1cm}p{1cm}p{0.8cm}}
\hline
 &  Material    &Supplier   &Technique& units&  $^{238}$U & $^{226}$Ra &    $^{232}$Th & $^{228}$Th &   $^{235}$U & $^{40}$K    & $^{60}$Co & $^{137}$Cs \\ \hline
1&  Pb& Britannia&  Ge& mBq/kg& &   $<$0.83    &&  $<$0.48    &&  $<$1.3  &$<$0.08&  \\
2&  Pb& Britannia&  GDMS&   mBq/kg& 0.35&&      0.094   &&&     0.12&&      \\
3&  Cu& Lugand Aciers&  Ge& mBq/kg& $<$4.1&    $<$0.16&   $<$0.15&   $<$0.13&   $<$0.17    &$<$0.37&  0.04$\pm$0.01   &$<$0.04\\
4&  Cu& Lugand Aciers&  GDMS&   mBq/kg& $<$0.012&&     $<$0.004   &&&     0.062   &&  \\
5&  316Ti SS, 10mm&    Nironit&    GDMS&   mBq/kg& $<$5.0 &&  $<$0.12&&&         $<$0.16    &&  \\
6&  316Ti SS, 15mm&    Nironit &GDMS&  mBq/kg& $<$9.9 &&  $<$0.41 &&&            $<$0.12    &&\\
7&  316Ti SS, 50mm&    Nironit &GDMS&  mBq/kg& $<$7.4 &&  $<$0.12 &&&            $<$0.09    &&\\
8&  Polyethylene&   In2Plastics &Ge&    mBq/kg& $<$18& $<$0.88&   $<$0.81&   $<$0.70&   $<$0.4 &$<$3.4&   $<$0.14&   $<$0.14\\
9&  Polyethylene&   Simona& ICPMS&  mBq/kg& $<$0.062&&     $<$0.021&&&&&              \\
10& Silver epoxy&   Circuit Works&  Ge& mBq/kg& $<$1.0 10$^{3}$&   13.6$\pm$2.8&   $<$18& $<$ 16& $<$4.5&    $<$52& $<$1.9&    $<$2.2\\
11  &Cu, rod&   Lumetalplastics &GDMS&  mBq/kg& 0.66$\pm$0.09& &        0.45$\pm$0.08   &&&     0.16    &&  \\
12  &Cu, sheet& Lumetalplastics &GDMS&  mBq/kg& 0.041$\pm$0.007 &&  0.014$\pm$0.002 &&&     0.031   &&  \\
13& Kapton-Cu PCB&  Flexiblecircuits&   Ge& mBq/unit&   $<$1.3 &0.031$\pm$0.004    &0.027$\pm$0.008&   0.042$\pm$0.004 &&  12.1$\pm$1.2&   $<$0.01&   $<$0.01\\
14& NTC sensors&    Murata  &Ge&    $\mu$Bq/unit&   $<$96& $<$1.5&    $<$1.6 &$<$1.3&   $<$0.3&    $<$2.9&    $<$0.2&    $<$0.2\\
15& LEDs&   Osram  & Ge&    $\mu$Bq/unit&   $<$90  &1.4$\pm$0.2    &3.5$\pm$0.4    &3.0$\pm$0.3&   $<$0.6&    $<$4.0&    $<$0.2 &$<$0.3\\
16& Plexiglas/PMMA&   Evonik& Ge& mBq/kg& $<$208 &$<$2.2&   $<$3.9 &$<$3.4&   $<$1.1 &$<$8.1&   $<$0.4 &$<$0.6\\
17& PMTs&   Hamamatsu   &Ge&    mBq/unit&   $<$87& $<$0.96&   $<$2.5 &0.69$\pm$0.35& 0.4$\pm$0.2 &11.5$\pm$2.1&  3.7$\pm$0.3 &$<$0.3\\
18& Cu  &Lugand Aciers  &GDMS&  mBq/kg& 0.025$\pm$0.005 &&  0.015$\pm$0.004    &&&     0.19    &&  \\
19& Capacitors& AVX &Ge&    $\mu$Bq/unit&   $<$360&    72$\pm$3    &749$\pm$3& 32$\pm$2&&      71$\pm$9&   $<$1&  $<$1\\
20& Resistors&  Finechem    &Ge&    $\mu$Bq/unit&   85$\pm$23&  4.1$\pm$0.3 &5.6$\pm$0.5&   4.4$\pm$0.3 &&  83.6$\pm$8.7&   $<$ 0.2 &104$\pm$11\\
21& Pin receptacles&    Farnell&    Ge& $\mu$Bq/unit&   217$\pm$42  &$<$1.1    &5.6$\pm$0.5    &4.5$\pm$0.4&   6.1$\pm$0.5 &20.5$\pm$2.4   &$<$0.3    &$<$0.2\\
22& Thermal epoxy&  Electrolube &Ge& mBq/kg&(1.0$\pm$0.2)10$^{3}$
&169.4$\pm$7.9& 52.1$\pm$3.7 &54.4$\pm$3.2   &&  105$\pm$12& $<$1.1&
$<$1.3\\ \hline

\end{tabular}
\caption{Activities measured in relevant materials for NEXT following different techniques. GDMS and ICPMS results were derived from U, Th and K concentrations. Germanium $\gamma$-ray spectrometry results reported for $^{238}$U and $^{232}$Th correspond to the upper part of the chains (derived from $^{234m}$Pa and $^{228}$Ac emissions) and those of $^{226}$Ra and $^{228}$Th give activities of the lower parts.}
\label{tr}
\end{table}
\end{landscape}
\normalsize

\end{document}